\documentclass[11pt,preprint]{aastex}

\newcommand{\Msun}{~M_\odot}

\newcommand{\ergs}{\rm ~erg~s^{-1}}

\tighten

\slugcomment{Submitted to ApJ; July 4, 2001}
\shorttitle{PWN IN EVOLVED SNRS}

\begin{document}

\title{PULSAR WIND NEBULAE IN EVOLVED SUPERNOVA REMNANTS}

\author{John M. Blondin}
\affil{Dept. of Physics, North Carolina State University, Raleigh, NC 27695}

\author{Roger A. Chevalier}
\affil{Dept. of Astronomy,
University of Virginia, P.O. Box 3818, Charlottesville, VA 22903}
\and
\author{Dargan M. Frierson}
\affil{Dept. of Mathematics, Princeton University, Princeton,  NJ 08544}

\begin{abstract}
For pulsars similar to the one in the Crab Nebula, most of the energy
input to the surrounding wind nebula occurs on a timescale $\la 10^3$
years; during this time, the nebula expands into freely expanding
supernova ejecta.
On a timescale $\sim 10^4$ years, the interaction of the supernova
with the surrounding medium drives a reverse shock front toward the
center of the remnant, where it crushes the PWN (pulsar wind nebula).
One- and two-dimensional, two-fluid simulations of the crushing and
re-expansion phases of a PWN show that (1) these phases are subject to
Rayleigh-Taylor instabilities that result in the mixing of thermal
and nonthermal fluids, and (2) asymmetries in the surrounding interstellar
medium give rise to asymmetries in the position of the PWN relative
to the pulsar and explosion site.
These effects are expected to be observable in the radio emission from
evolved PWN because of the long lifetimes of radio emitting electrons.
The scenario can explain the chaotic and asymmetric appearance of
the Vela~X PWN relative to the Vela pulsar without recourse to a
directed flow from the vicinity of the pulsar.
The displacement of the radio nebulae in  G327.1--1.1,
MSH15--56 (G326.3--1.8), G0.9+0.1, and W44 relative to the X-ray 
nebulae may be due to this mechanism.
On timescales much greater than the nebular crushing time, the
initial PWN may be mixed with thermal gas and become unobservable,
so that even the radio emission is dominated by recently injected
particles.
\end{abstract}


\section{INTRODUCTION}

Pulsars are expected to be born inside massive stars, so that the
evolution of the wind nebulae that they produce is expected to depend
on a number of different factors: the structure of the supernova,
the nature of the surrounding medium, the evolution of the pulsar
spin-down power, and the space velocity of the pulsar.
The evolution of pulsar nebulae can be divided into 
a number of phases that are 
important for their observational appearance
(Reynolds \& Chevalier 1984, hereafter RC84; Chevalier 1998). 
Initially, the pulsar nebula expands into the freely expanding
supernova ejecta.
The pulsar provides a constant wind power and the
swept-up shell of ejecta is accelerated.
Jun (1998) has carried out two-dimensional simulations of
the Rayleigh-Taylor instabilities that occur during this phase, with
the aim of modeling the Crab Nebula.
On a timescale of $\sim$ $10^3$ yr for a pulsar like the Crab, the power 
input from the pulsar drops steeply so that the nebula expands adiabatically.
The expansion approaches free expansion within the supernova.  

The next phase of evolution results from the interaction of the
supernova remnant with the surrounding medium and the inward motion
of the reverse shock front that is driven by this interaction.
On a timescale of $\sim$ $10^4$  yr,  the reverse shock makes 
its way back to the center of the remnant.  
The pulsar nebula that has been created by the early energy injection
is compressed during this phase.
Because of the synchrotron losses of high energy particles, this nebula
is best observed at radio wavelengths.
At the same time, the continued wind power from the pulsar can
create a nebula, including high energy emission, that is
localized to the pulsar.
If the pulsar has a velocity of 100's of km s$^{-1}$, as 
is quite likely for a normal pulsar (Lyne \& Lorimer 1994), 
a bow shock nebula can 
form around the pulsar and this nebula can separate from the
crushed nebula from the earlier phase.

RC84 made some approximate estimates of
the effects of crushing a pulsar nebula by the external supernova
remnant, with an emphasis on the synchrotron luminosity.
Van der Swaluw et al. (2000) carried out one-fluid, one-dimensional simulations
of the crushing process, finding that there are considerable
transient effects before the nebula settles into a slow expansion.
Our aim here is also to investigate the hydrodynamics of the interaction,
with attention to instabilities.
In \S~2, we present the basic parameters that are needed to model
the pulsar nebula/supernova remnant interaction.
A difference with the work of van der Swaluw et al. (2000) is that
they assumed supernova ejecta with a constant density profile,
whereas we allow for an outer power law profile.
Numerical simulations in one and two dimensions are presented
in \S~3 and \S~4, respectively.
We allow for different adiabatic indices in the relativistic
fluid and the thermal gas fluid.
In the two-dimensional simulations, 
the effect of a density gradient in the external medium is treated
in addition to a constant density ambient medium.
We believe this situation is relevant to the Vela X radio pulsar
nebula in the Vela supernova remnant, which is discussed in \S~5
along with other remnants.
The conclusions  are in \S~6.

\section{A MODEL FOR THE PWN/SNR INTERACTION}

To investigate the dynamical evolution of the pulsar wind nebula/supernova
remnant system beyond the free expansion phase, we have constructed a simple
model based on the expanding pulsar bubble solution (Chevalier 1977) and
the self-similar driven wave (Chevalier 1982).  We begin by assuming the
stellar material ejected by the Type II supernova is expanding ballistically
($r = vt$) and can be described by a steep outer power law density
profile inside of which the density is constant.
The density profile is parameterized by the total mass of ejecta, $M_{ej}$,
the kinetic energy released in the supernova, $E_{sn}$, and the density
power law exponent, $n$:
\begin{equation}
\rho_{ej} (r,t) = \left\{
\begin{array}{l@{\quad {\rm for} \quad}l}
A v_{t}^n r^{-n} t^{n-3} & r > v_t t \\
A t^{-3} & r < v_t t \\
\end{array} \right.
\end{equation}
where the constant $A$ is given by
\begin{equation}
A = \frac{5n-25}{2\pi n} E_{sn} v_t^{-5}
\end{equation}
and the velocity at the intersection of the density plateau and the power law
is given by
\begin{equation}
v_t = \left(\frac{10n-50}{3n-9}\frac{E_{sn}}{M_{ej}}\right)^{1/2}.
\end{equation}

If we further assume that this ejecta is expanding into a 
uniform ambient medium with density $\rho_a$, 
the expansion of the supernova remnant is 
described by a self-similar solution given by Chevalier (1982).  The radius
of the forward shock, $R_1$, in this 
self-similar driven wave (SSDW) is given by (assuming $s=0$ in
Chevalier's notation)
\begin{equation}
R_1 = \alpha \left(\frac{\rho_t v_t^n}{\rho_a}\right)^{1/n} 
t^{(n-3)/n}
\end{equation}
where the constant $\alpha = 1.048$ for $n=9$, but varies relatively little
with the value of $n$.
The assumption of a constant density surrounding medium may not 
generally apply because core collapse supernovae have massive
star progenitors which are known to affect their surroundings
through winds and photoionization.
The timescale for the reverse shock wave to return to the center
($\sim 10^4$ years) makes it plausible that the outer shock wave
has proceeded to the interstellar medium.
If a wind bubble has been created by the progenitor star, the
inward motion of the reverse shock is modified.

The last component of the model, the pulsar wind nebula, is treated as
an adiabatic wind bubble driven by a pulsar wind with a kinetic
luminosity, $L_p$.  If we use the standard approximation of an
isobaric bubble bounded by a thin shell of swept up ambient gas, 
as in Chevalier (1977),
the expansion of this bubble within the uniform
(but decreasing with time) density of the ejecta is given by
\begin{equation}
R_p = 1.50\left(\frac{n}{n-5}\right)^{1/5}
\left(\frac{n-5}{n-3}\right)^{1/2}
\left(\frac{E_{sn}^3 L_p^2}{M_{ej}^5}\right)^{1/10} t^{6/5}.
\label{earlyrp}
\end{equation}

\begin{figure}[!hbtp]
\plotone{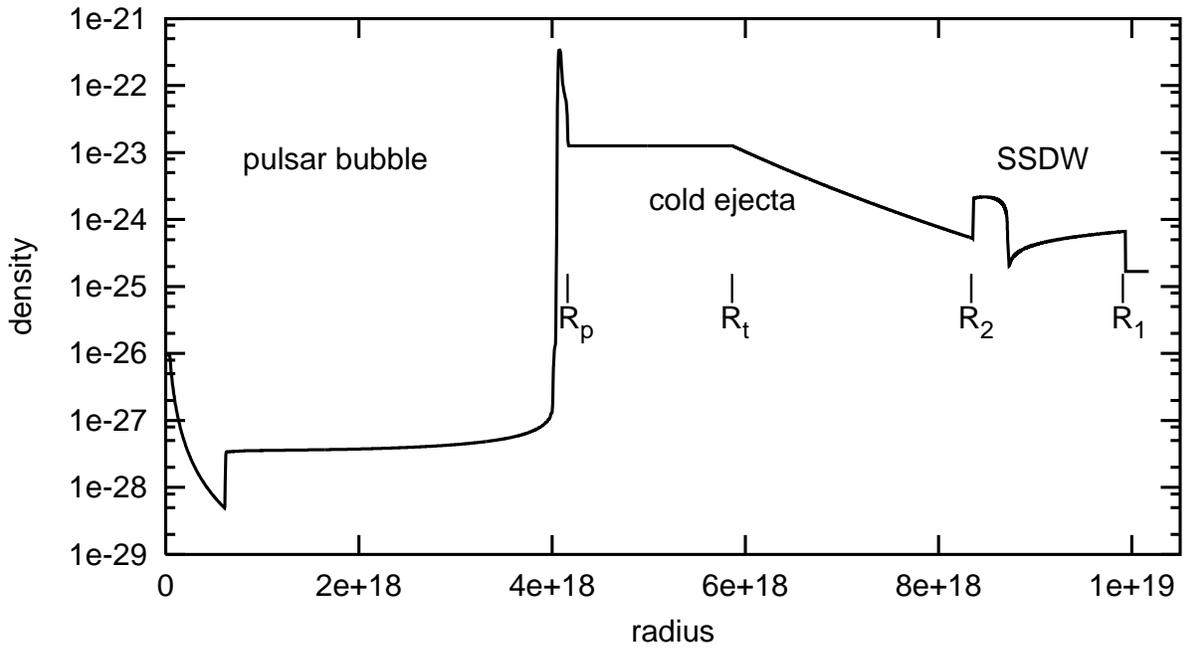}
\figcaption{The density profile of our model for the interaction of
a pulsar nebula with the host supernova remnant.  The supernova remnant
is modeled as a self-similar driven wave (SSDW) bounded by a forward
shock at $R_1$ and a reverse shock at $R_2$.  The pulsar bubble has 
swept up a thin shell of ejecta at $R_p$.  The edge of the ejecta
plateau at $R_t$ is just about to reach $R_2$, after which the
reverse shock will begin propagating in toward the center.
\label{fig:bif}}
\end{figure}

This combined model is illustrated in Figure \ref{fig:bif} which shows the
radial profile of the density throughout the structure.  The supernova 
ejecta is separated into three parts: a thin shell at $R_p$ that 
has been swept up by the expanding pulsar bubble, cold freely expanding
ejecta between $R_p$ and the reverse SNR shock at $R_2$, and a thin
shell of shocked ejecta at $R_2$ that has been 
decelerated by the reverse shock. 
The discontinuity between
$R_2$ and $R_1$ separates the shocked ejecta (on the left) and
the shocked CSM (on the right).  The discontinuity near $r=0$ is
the termination shock of the pulsar wind.

Several things will happen at an age of $\sim 10^3$ years to 
change this model.  First, the pulsar will die out.  Without
continued power input, the expansion of the pulsar nebula will slow down
until it becomes frozen into the expanding ejecta.  Second, the pulsar
nebula will reach the edge of the plateau in the SN ejecta and begin 
accelerating down the steep density gradient.  Alternatively, the plateau  will
reach the reverse shock of the SSDW.  As the ram 
pressure at the reverse shock rapidly decays away, the reverse shock
is driven toward the center of the remnant, resulting in a dynamical
interaction with the pulsar nebula.  

The exact ordering of these events depends on the parameters of the
model described above, namely $E_{sn}$, $M_{ej}$, $n$, $\rho_a$, $L_p$
and the lifetime of the pulsar.  The time of the first
event, the ``death'' of the pulsar, is determined by our model for
$L_p(t)$. 
We approximate the power input from the pulsar by 
assuming a constant pulsar magnetic field and
braking index $p$, which yields
\begin{equation}
L_p = L_{pi}\left(1+{t\over \tau}\right)^{-(p+1)/(p-1)},
\label{lumevol}
\end{equation}
where $L_{pi}$ is the initial pulsar power and $p=3$ for the
magnetic dipole case.
It can be seen that much of the power input occurs up to time $\tau$,
so that $\tau$ corresponds roughly to the lifetime of the pulsar.
The initial pulsar power is given in terms of the initial spin period,
$P_i$: 
$L_{pi}=I\Omega_i^2/[(p-1)\tau]$, where $I$ is the pulsar
moment of inertia and $\Omega_i=2\pi/P_i$ is the initial spin rate.
For the Crab pulsar, $p$ is measured to be 2.5 and $L_{p}=4\times 10^{38}\ergs$,
which lead to $\tau=744$ years
and $L_{pi}=3\times 10^{39}\ergs$
for $I=10^{45}$ g cm$^{2}$
(e.g., Chevalier 1977 and references therein).
The power input for our model C (see Table 1) is considerably above $L_{pi}$ for
the Crab pulsar, but corresponds to $P_i\approx 5$ ms (for $p=3$), which
is a plausible initial rotation period.

The second event, the pulsar nebula reaching the edge of the 
ejecta plateau, $R_t = v_t * t$, occurs at
\begin{equation}
t_2 = 2.64 \left(\frac{n-5}{n}\right)\frac{E_{sn}}{L_p}
\end{equation}
assuming constant pulsar luminosity.  If, however, the pulsar has
already died away, the nebula will expand slower and this will be an 
underestimate to $t_2$.
The third event, the inward motion of the reverse shock, begins when the
edge of the ejecta plateau reaches the reverse shock, i.e., $R_t=R_2$:
\begin{equation}
t_3 =  B\left(\frac{\rho_t}{\rho_a}\right)^{1/3}\approx
120 \left(\frac{M_{ej}}{\Msun}\right)^{5/6}
    \left(\frac{\rho_a}{m_p}\right)^{-1/3}
    \left(\frac{E_{sn}}{10^{51} {\rm erg}}\right)^{-1/2} {\rm yrs},
\end{equation}
where the numerical value is quoted for $n=9$, for which 
$B=0.688$.

Choosing reasonable parameters as listed in Table \ref{tabl:params} for Model A, 
we find $t_2 \approx 3,700$ yrs and $t_3 \approx 1,500$ yrs.  Thus
we expect the reverse shock to begin moving into the interior of
the remnant where it will encounter the pulsar nebula. 
For other parameters, however, this sequence may be reversed.  For
example, lowering $E_{sn}$ by 3 would lead to $t_2 < t_3$; 
the lower energy means a slower expansion, requiring a longer time
to sweep up mass but a shorter time for the pulsar to blow through
the slower ejecta.

\begin{table}
\begin{tabular}{|l l l r r r r r|}
\hline
Parameter & Definition & Units & Model: A  &  B  &  C & D & E\\
\hline
$L_{pi}$   & pulsar luminosity    & $10^{40}$ erg s$^{-1}$ &  1.0 & 0.1 & 5.0 & 5.0 & 1.0 \\
$\tau$& pulsar lifetime      & yrs             & 500 & 500 & 500 & 500 & 2500\\
$M_{ej}$    & mass of SN ejecta    & M$_\odot$       & 8 & 4 & 16 & 8 & 8\\
$E_{sn}$    & kinetic energy of SN & $10^{51}$ erg   & 1 & 1 & 1 & 1 & 1    \\
$n$         & ejecta power law     &                 & 9 & 9 & 9 & 9 & 9 \\
$\rho_{a}$  & ambient density      & $m_p$ cm$^{-3}$ & 0.1 & 0.1 & 0.1 & 0.1 & 0.1\\
\hline
\end{tabular}
\label{tabl:params}
\end{table}

\section{ONE-DIMENSIONAL HYDRODYNAMIC SIMULATIONS}

To investigate the dynamical evolution of the pulsar nebula beyond the 
phase of self-similar expansion in a uniform ejecta,
we employ numerical simulations that begin shortly before the PWN/SNR interaction. 
These simulations are computed with the
VH-1 code, a Lagrangian-Remap version of the Piecewise Parabolic Method.
The standard VH-1 code was adapted to treat two fluids with two different
values of $\gamma$ such that the pulsar wind can be modeled with $\gamma = 4/3$
while the supernova ejecta and circumstellar gas are modeled with $\gamma = 5/3$.
The one-dimensional PWN/SNR model is evolved on a grid of 1000 radial zones 
that expands to follow the forward shock.  Several simulations on a grids
of 750 and 1500 zones produced almost identical results.  

The initial conditions for the SSDW are taken from the solution
described in Chevalier (1982), scaled to the parameters listed in Table
\ref{tabl:params} for Model A.  The initial conditions for the pulsar nebula
are taken from a separate hydrodynamic simulation of a wind blown
bubble in which the wind is an adiabatic gas with $\gamma = 4/3$, a
mass loss rate $\dot M_w = 2L_p/v_p^2$, and a wind velocity,
$v_p$, scaled to yield a sound speed in the shocked gas of $c/\sqrt
3$, appropriate for a relativistic gas.  The pulsar luminosity decays
away according to eq. (\ref{lumevol}) assuming a braking index of $p=3$.

A more realistic model for the pulsar wind nebula would allow for
a relativistic, MHD (magnetohydrodynamic) flow, as in the steady state
model for the Crab Nebula by Kennel \& Coroniti (1984a,b).
However, once the relativistic wind passes through the wind termination
shock, the flow velocity drops to $c/3$. 
The bulk velocity is not relativistic in the nebula, although the
shocked particles are highly relativistic.
This situation can be approximated by a $\gamma = 4/3$ fluid.
The magnetic field may be a more significant omission because
the magnetic field in the Crab Nebula is thought to be in
approximate equipartition with the particles.
In particular, the magnetic field could affect the growth of
instabilities at the edge of the pulsar nebula (Jun, Norman, \& Stone 1995;
Hester et al. 1996).
The tangential magnetic field may be especially strong at the
edge of the nebula (Emmering \& Chevalier 1987). 
The  magnetic field may  inhibit the growth of the instability along
the direction of the field, but has less of an effect across the field.

\begin{figure}[!hbtp]
\plotone{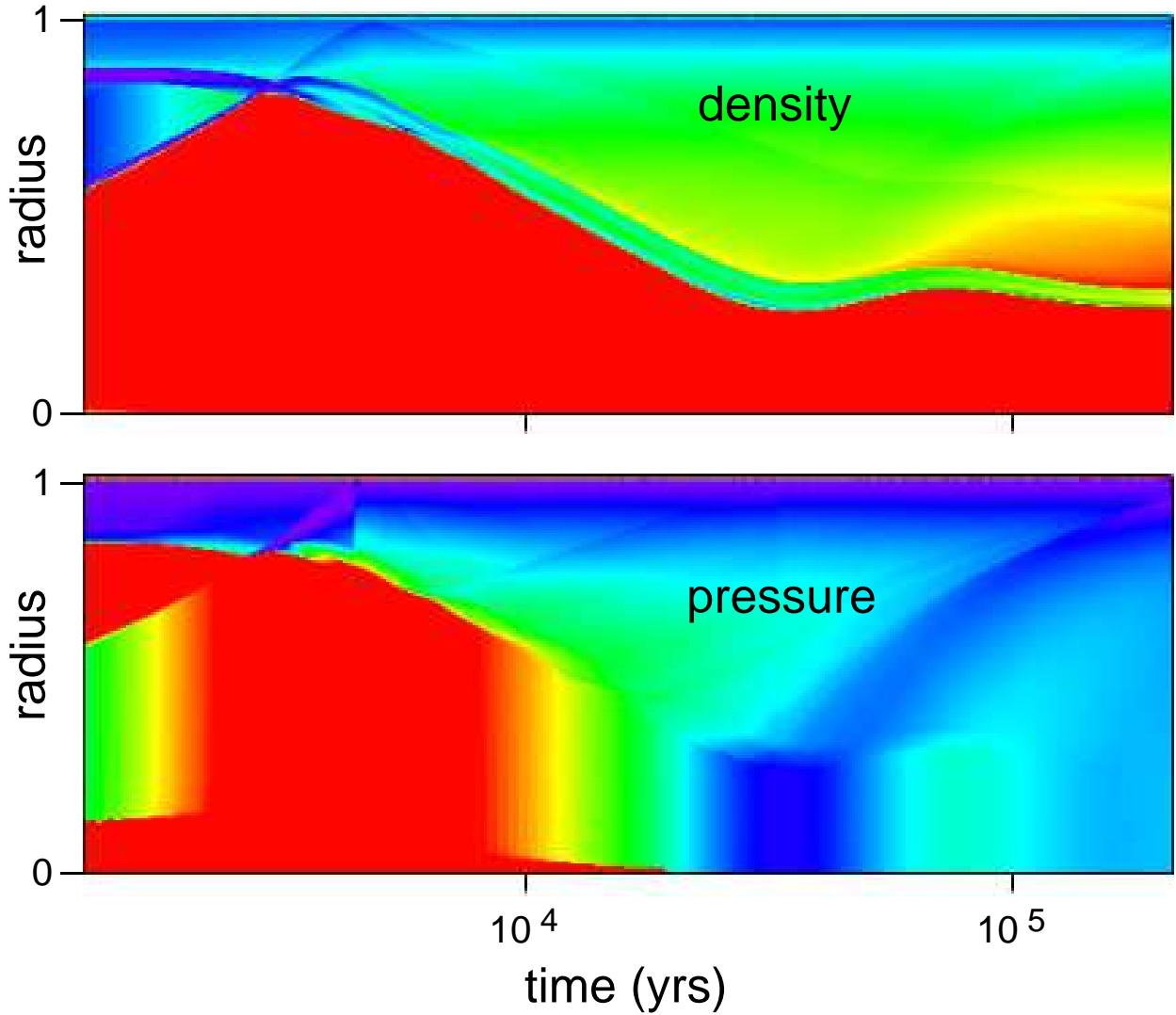}
\figcaption{Evolution of the pulsar nebula within the host supernova remnant
as computed with the one-dimensional hydrodynamics model. Both the density
(top) and the pressure (bottom) are depicted in logarithmic scales, with
yellow corresponding to high values and purple to low values.
The radial extent
of the grid is scaled to follow the forward shock front, such that the location
of the forward shock is a horizontal line near the top of each plot. 
\label{fig:pnsnb}}
\end{figure}

The evolution of the PWN/SNR model is illustrated with a space-time
diagram in Figure \ref{fig:pnsnb}.  
The evolution begins at $t\approx 2\tau$, just
prior to the end of the SSDW phase at $t_3$.
One can identify two bands of high-density gas
in the early evolution: the pulsar
nebula shell made up of swept up ejecta, and a shell of shocked
ejecta on the trailing side of the SSDW.  The simulation begins with
the edge of the ejecta plateau reaching the reverse shock.  The reverse
shock and the shell of shocked ejecta then begin to push inward,
but the SSDW does not have much time to react
to the dropping ejecta density before the pulsar shell strikes it.

At an age of about $\sim 3000$ years the pulsar nebula shell collides
with the dense SSDW shell.  From this point on all of the ejecta is
contained in a relatively thin shell (width $\sim 0.1\, R_1$) bounded
by low density shocked pulsar wind on the inside and by shocked
circumstellar gas on the outside.  This initial impact generates a strong
shockwave transmitted through the SSDW.  When this shock catches up with
the forward shock there is an abrupt, but brief, acceleration of the 
forward shock, followed by a faster deceleration than in the SSDW phase.
A weak reflected shock is also generated, which bounces back and forth
through the pulsar bubble at the relatively high sound speed of $c/\sqrt 3$.
This bouncing wave can be seen as nearly vertical lines in the pressure image
of Figure \ref{fig:pnsnb}.

Following the collision of the pulsar and reverse SNR shocks, 
the thin shell of ejecta is
gradually pushed inward with respect to the outer shockwave by the 
high pressure of the shocked CSM.  Initially the pressure in the
pulsar bubble is negligible, and the presence of the PWN has no
effect on the dynamics of the crashing reverse shock.  But as the pulsar
bubble is compressed, the interior pressure rises.
For our nominal
PWN/SNR parameters, the pressure in the bubble becomes comparable to
the pressure behind the reverse shock once the PWN has been crushed to
approximately half the radius of the SNR.  Eventually the PWN 
reaches a minimum radius of
$0.25\, R_1$ at an age of $\sim 35,000$ years.  At this point the
pressure of the compressed pulsar bubble accelerates the shell of
shocked ejecta back out to a radius of $0.3\, R_1$.  From this point on
all of the ejecta remains in a thin shell with a width of $\sim 10\%$ of $R_1$,
slowly oscillating about an average radius of $\sim 0.27\, R_1$.
The simulations of van der Swaluw et al. (2000) also show oscillations
after the crushing phase.

The extent to which the pulsar bubble is compressed depends on the pressure
in the bubble compared to the pressure behind the reverse shock as it propagates
in toward the pulsar nebula.  Assuming a constant pulsar luminosity
and using the pressure behind the reverse shock
in the SSDW phase, we find a ratio of pressures at a time of $t_3$:
\begin{equation}
\frac{P_b}{P_2} = 0.166\,  L_p^{2/5}\,  M_{ej}^{1/3}\,
           \rho_a^{-2/15}\, E_{sn}^{-3/5},
\label{eqn:pratio}
\end{equation}
assuming an ejecta exponent of $n=9$.
This ratio is $\sim 0.2$ for our nominal parameters.  Note that both of
these pressures will be slightly lower: $P_b$ because the pulsar has likely
died out by the time of the collision, $P_2$ because the pressure in the
driven wave will drop once the reverse shock starts falling to the center.

\begin{figure}[!hbtp]
\epsscale{0.7}
\plotone{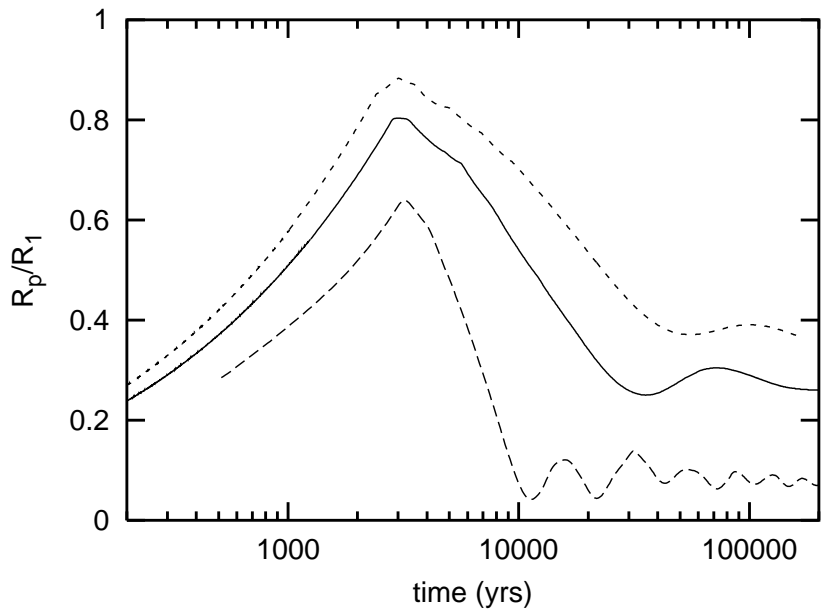}
\figcaption{Evolution of the pulsar nebula through the return of the
reverse SNR shock for the three different models listed in Table 1:
Model A (solid), Model B (long dash), and Model C (short dash).
The radius of the pulsar nebula is scaled to the
(increasing) radius of the SNR as in the previous figure.
\label{fig:rpn1}}
\end{figure}

We have run additional models  in order to show the effects of changing 
the energy in the pulsar bubble on the evolution of the PWN/SNR interaction.
In Model B we dropped the pressure ratio in eq. (\ref{eqn:pratio}) by
a factor of $\sim 3$ through a smaller ejecta mass ($4\Msun$) and smaller 
pulsar luminosity ($10^{39}\ergs$).  In Model C the 
pressure ratio was increased by a factor of 2.4 by doubling the ejecta
mass ($16\Msun$) and increasing the pulsar luminosity 
($5\times 10^{40}\ergs$).  The evolution of all three models is 
shown in Figure \ref{fig:rpn1}.  As expected, the lower the pressure in
the pulsar bubble, the more the bubble is compressed by the reverse shock.
In the lower pressure model the radius of the bubble after collapse is
only $\sim 0.06\, R_1$.

The evolution in the high-pressure model is slightly different since
these parameters switch the ordering of $t_2$ and $t_3$.  In this case
the pulsar nebula hits the edge of the ejecta plateau at $\sim \tau$,
well before the PWN/SNR collision.  There is thus an intermediate phase
when the pulsar shell accelerates down the steep ejecta gradient before
impacting the SSDW.  This results in a more violent collision, which 
compresses the SSDW by a factor of $\sim 2$ and accelerates the forward
shock considerably more than in the standard model.  Despite this more violent
beginning, however, the subsequent evolution is very similar to the 
standard model, albeit leaving behind a larger pulsar bubble.

It can be seen in Figure \ref{fig:rpn1} that after the PWN has been crushed, the
ratio $R_p/R_1$ evolves slowly with time, 
if the pressure wave induced oscillations
are averaged out.
If approximate pressure equilibrium is established between the
PWN and the interior of the blast wave, we expect $R_p\propto t^{0.3}$ (RC84).
The radius of a Sedov blast wave evolves as 
\begin{equation}
R_1=\left(2.025 E_{sn}\over \rho_a\right)^{0.2} t^{0.4},
\label{sedov}
\end{equation}
 so the slow
evolution of $R_p/R_1$ is expected.
Van der Swaluw \& Wu (2001) have emphasized this point, noting that
equating the pressure in the PWN with the interior pressure in
the blast wave leads to
\begin{equation}
{R_p\over R_{1}}=1.02\left(E_{pwn}(t)\over E_{sn}\right)^{1/3},
\label{rp}
\end{equation}
where $E_{pwn}(t)$ is the internal energy in the PWN at time $t$.
We now assume that the pulsar power can be assumed to be constant
up to time $t_p$, after which it falls to 0.
The value of $E_{pwn}(t)$ can be related to the total spindown
energy of the pulsar $E_{sd}=L_{pi} t_p$ for $p=3$.
During the initial expansion phase ($t\le t_p$), the radius of the PWN is 
given by eq. (\ref{earlyrp}) and $E_{pwn}(t)=5L_p t/11$ (RC84).
For $t>t_p$, the evolution of the PWN is adiabatic, so
that $E_{pwn}(t)=E_{pwn}(t_p)R_p(t_p)/R_p(t)$, where
$E_{pwn}(t_p)=5E_{sd}/11$.
With the use of eqs. (\ref{earlyrp}) and (\ref{sedov}),
eq. (\ref{rp}) can be expressed as
\begin{equation}
{R_p\over R_{1}}=0.866\left(E_{sd}\over E_{sn}\right)^{0.3}
\left[\left(n-5 \over n-3\right)^{0.5}\left(n\over n-5\right)^{0.2} 
{E_{sn}^{0.3}\rho_a^{0.2} t_p\over M_{ej}^{0.5}}\right]^{0.25} t^{-0.1}.
\label{rpfinal}
\end{equation}
Application of eq. (\ref{rpfinal}) to the 3 models discussed here
yields $R_p/ R_{1}= 0.20$ (Model A), 0.11 (Model B), and 0.29 (Model C)
at $t=5\times 10^{12}$ s if we take $t_p=\tau$.
It can be seen from Fig. (3) that these values are slightly lower than
seen in the simulations.  This discrepancy is due in large part to the
time-dependence of $L_p$, such that some pulsar energy continues to be
added to the bubble after a time $\tau$.  In simulations with a constant
value of $L_p$ up until $t_p$ and zero thereafter, the radius of the
crushed PWN at late times was better fit by this approximation.

We have $E_{sd}=I\Omega_i^2/2$, so that the observed value
of $R_p/ R_{1}$ can be related to the initial spin rate of the
pulsar (van der Swaluw \& Wu 2001); since $R_p/ R_{1}\propto \Omega_i^{0.6}$,
the value of $R_p/ R_{1}$ appears to be more sensitive to $\Omega_i$
than to the other uncertain parameters.
However, there is some sensitivity to the power input timescale, $\tau$,
because if $\tau$ is small, the wind power is injected at
an early time and the bubble energy is more subject to adiabatic
expansion losses.
In the magnetic dipole model with braking index $p=3$, $\tau$ in
eq. (\ref{lumevol}) is given by (e.g., Shapiro \& Teukolsky 1983)
\begin{equation}
\tau = {3Ic^3 \over \Omega_i^2 B_{ns}^2 R_{ns}^6 \sin^2\alpha},
\label{tau-p}
\end{equation}
where $B_{ns}$ is the magnetic dipole field strength of the pulsar,
$R_{ns}$ is its radius, and $\alpha$ is the angle between the dipole
field and the rotation axis of the pulsar.
Ordinary radio pulsars appear to have magnetic field strengths
in a fairly narrow range around $ 3\times 10^{12}$ G (e.g., Stollman 1987),
comparable to the field strength of the Crab pulsar.
There may be a separate class of strongly magnetized neutron stars,
magnetars, with $B_{ns}\ga 10^{14}$ G (Duncan \& Thompson 1992);
these objects are expected to rapidly deposit their rotational
energy into the surroundings.
Turning to the initial spin rate and taking $\tau\propto \Omega_i^{-2}$
as given by eq. (\ref{tau-p}), we have $R_p/ R_{1}\propto \Omega_i^{0.1}$.
The insensitivity of $R_p/ R_{1}$ to $\Omega_i$ is due to the fact that
a pulsar with a large spin down energy tends to deposit the energy
into the nebula early, when it is subject to adiabatic expansion losses.
In addition, early radiative losses can decrease the energy in the
pulsar bubble (see also van der Swaluw \& Wu 2001).
The radiative luminosity of the Crab Nebula at present is $\sim 1/3$
of the spin-down power (Davidson  \& Fesen 1985), and the higher magnetic
field at earlier times can make the losses even more important.
Atoyan (1999) has described a model for the Crab Nebula in which the
initial period is $\sim 4$ msec and most of the spin down energy is lost to
radiation.
For these combined reasons, the observed value of $R_p/ R_{1}$ 
for an evolved pulsar wind nebula is
unlikely to lead to a reliable estimate for the initial spin rate.

\begin{figure}[!hbtp]
\epsscale{0.7}\plotone{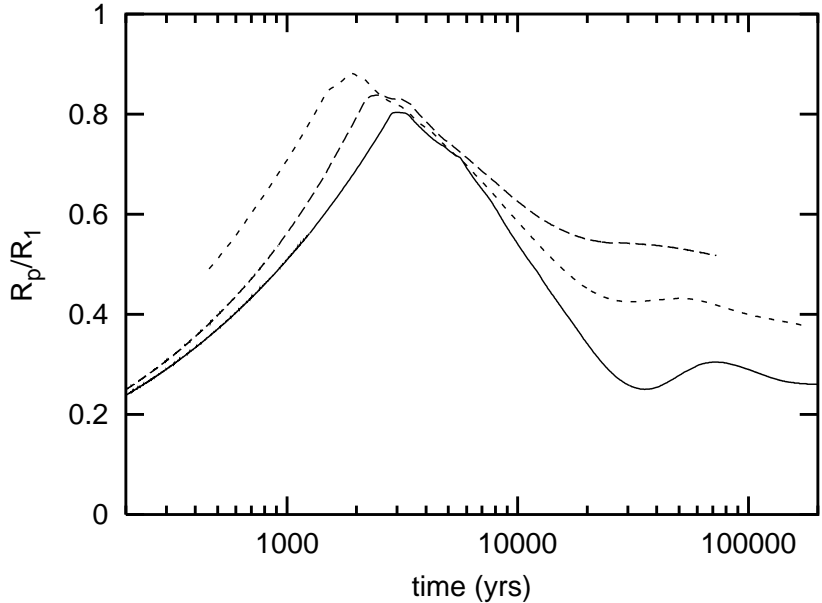}
\figcaption{Evolution of the PWN for models A, D, and E listed in 
Table \ref{tabl:params}.
The radius of the pulsar nebula is scaled to the
(increasing) radius of the SNR as in the previous figure.
\label{fig:rpn2}}
\end{figure}

To illustrate these effects we show, in Figure \ref{fig:rpn2}, 
the evolution of $R_p/ R_{1}$ for
three models with the same SNR parameters, but different pulsar winds.  In 
particular, we include two models with the same total spin down energy:
Model D with $L_p=5\times 10^{40}\ergs$ and $\tau_p=500$ yrs, and Model E 
with $L_p=10^{40}\ergs$ and $\tau_p=2500$ yrs.  Eq. (\ref{rpfinal})
predicts that the pulsar with a more gradual power output will produce 
a PWN at late times that is $50\%$ larger than a PWN from a pulsar with
the same spin down energy, but releasing it away 5 times faster.  

In addition to the PWN, radiative cooling can be important for the
outer SNR, resulting in the formation of a dense shell (Chevalier 1974;
Cioffi, McKee, \& Bertschinger 1988).
The shell radius expands approximately as $R_1\propto t^{0.3}$.
If both the PWN and the hot interior SNR gas evolve adiabatically
and $R_p^3 \ll R_1^3$, pressure equilibrium between the PWN and the SNR
implies that $R_p^4\propto R_s^5$ or $R_p/ R_{1}\propto t^{0.075}$.
The size of the PWN nebula relative to the SNR again remains approximately
constant with time, although now the PWN expands slightly more rapidly.
The hot SNR gas may not evolve adiabatically but may be subject to
radiative cooling just inside the dense shell (Chevalier 1974).
This has the effect of further increasing the ratio
$R_p/ R_{1}$.

\section{TWO-DIMENSIONAL HYDRODYNAMIC SIMULATIONS}

\subsection{Evolution in a Uniform Medium}

To investigate the effects of  instabilities on the morphology of the
pulsar nebula, we have extended our models to two dimensions.  We
computed these models on a cylindrical ($R$-$z$) grid of 2000 $\times$
2000 zones.  We assume reflection symmetry about the equator and
compute only one hemisphere.  The initial conditions are taken from the
first stage of the one-dimensional simulations just after the drop of
the pulsar wind termination shock.  In order to make these simulations
more computationally efficient, we have raised the density in the
pulsar bubble by a factor of 30.  This artificially lowers the sound
speed in the hot bubble allowing the use of a larger time step.
Although this increases the time needed to equilibrate the pressure in
the bubble, the sound crossing time is still relatively short compared
to the expansion time, and one-dimensional simulations with and without
this modification produced identical global evolutions.  Again, we
evolve this model on a moving grid that follows the outer shock front,
so that the full resolution of the grid is always used.

\begin{figure}[!hbtp]
\epsscale{1.0}
\plotone{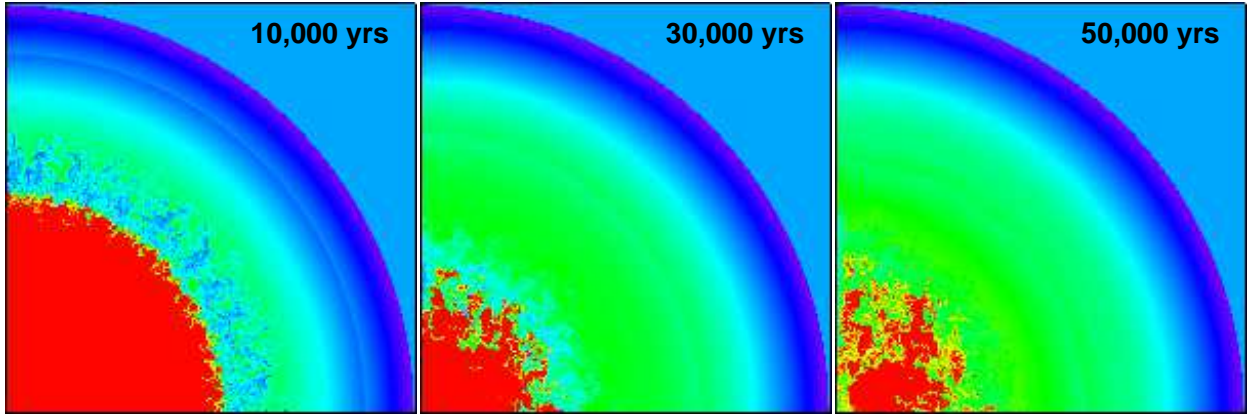}
\figcaption{Crushing of the pulsar nebula in a two-dimensional
simulation using the parameters for Model A listed in Table \ref{tabl:params}.
The images show the gas density on a logarithmic scale covering 2.5
orders of magnitude. The length scale of the images are normalized to 
take out the expansion of the outer shock front.
\label{fig:darg1}}
\end{figure}

The outer shock front remains very nearly spherical and follows the time
evolution of the one dimensional model to within $0.1\%$.  The key difference 
is the instability of the thin shell of ejecta and
the rapid mixing driven by this instability.  
Even before the PWN/SNR interaction,
the instability of the SSDW (Chevalier et al. 1992)
begins to spread out the shell of shocked ejecta.
However, given the short time between the beginning of the simulation 
and the crash of the reverse shock, 
there is not enough time for this instability to grow to significant amplitude.  
Following the PWN/SNR collision, the deceleration of the ejecta shell by
the shocked ambient medium
(responsible for the SSDW instability) increases in magnitude,
thereby driving a more rapid Rayleigh-Taylor instability.  The result is 
a broadened shell of mixed ejecta and ambient medium 
as seen in the first two frames of Figure \ref{fig:darg1}.  

Much more dramatic, however, is the Rayleigh-Taylor instability working
in the opposite direction when the high pressure of the compressed
pulsar bubble begins to accelerate the ejecta shell back outwards.  From
the 1D simulation shown in Figure \ref{fig:pnsnb}, we see that once 
$R_p/R_1$ drops below $\sim 0.5$, the pressure in the compressed PWN
exceeds that in the SNR and the PWN begins to decelerate the shell of
shocked ejecta.  This acceleration of the dense ejecta gas by the
low-density PWN gas is subject to the Rayleigh-Taylor instability.  As
a result of this instability, much of the ejecta gas continues toward
the center of the SNR almost unabated.  In the absence of a PWN, a spherical
reverse shock would reach the center of the remnant at an age of $\sim 35,000$ yrs,
and indeed we see in Figure  \ref{fig:darg1} that some has reached the
center by this time.  In the same process some relativistic gas is 
displaced from the center.

\begin{figure}[!hbtp]
\epsscale{0.7}
\plotone{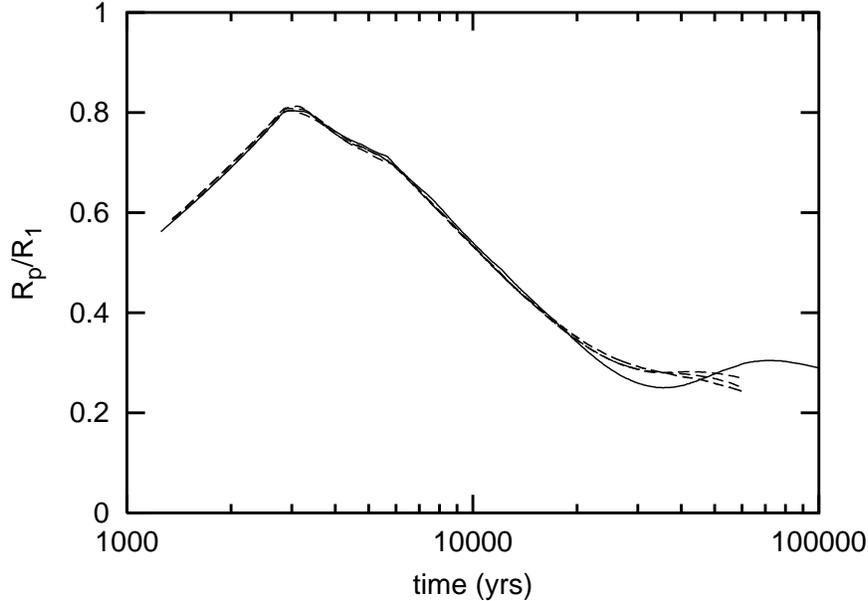}
\figcaption{Evolution of the pulsar nebula through the crash of the
reverse SNR shock in one (solid line) and two (dashed lines) dimensions
with different numerical resolutions.
\label{fig:rpn2d}}
\end{figure}

We see two competing effects of this instability in the PWN crushing phase.
First, as relativistic gas is displaced from the center, it escapes the
full compression of the infalling ejecta seen in 1D.  Second, the vigorous
turbulence driven by this instability leads to rapid mixing of the thermal
and relativistic gasses.  To illustrate these effects, we compare the
effective value of $R_p/R_1$ in the 1D and 2D simulations in 
Figure \ref{fig:rpn2d}.  We compute the radius of
the pulsar bubble in the 2D simulation by summing up the volume of
gas with $\gamma < 1.66$ (i.e., including the partially mixed gas) and
calculate an effective radius assuming a spherical volume.  Prior to
the bounce, the 1D and 2D simulations are nearly identical.  However,
at the moment of bounce, the PWN in the 2D simulation has a volume twice
that of the PWN in the 1D simulation.  The compressed PWN quickly rebounds 
in the 1D simulation, but in 2D the volume of relativistic gas continues
to shrink due to numerical mixing. 

\begin{figure}[!hbtp]
\epsscale{0.9}
\plotone{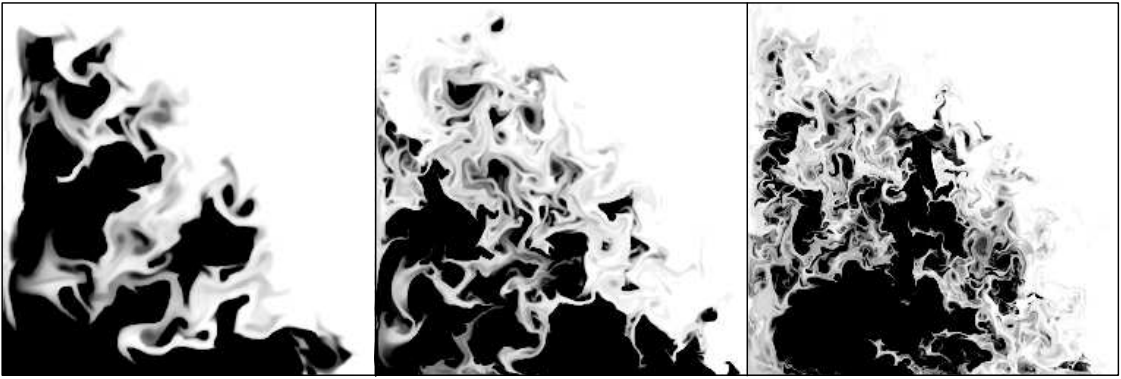}
\figcaption{The rapid mixing of the thermal gas of the SN ejecta (white) 
and the relativistic gas of the PWN (black) is seen in
these images of $\gamma$ at a time of 50,000 yrs, shortly
after the bounce of the pulsar bubble in the corresponding one-dimensional
simulation.  The three images correspond to increasing numerical resolution,
with the number of zones per dimension, $N = 500$ (left), 1000 (center),
2000 (right).
\label{fig:mixing}}
\end{figure}

This rapid depletion of the volume of the PWN in the 2D simulation is an artifact
of numerical diffusion; when one numerical zone contains both relativistic
gas ($\gamma = 4/3$) and ejecta gas ($\gamma = 5/3$), the mass-weighted
average $\gamma$ is dominated by the high density of the ejecta.  As a result,
the mixing - and subsequent loss of PWN volume - is strongly dependent on
the numerical resolution of the simulation.  Higher spatial resolution leads
to less numerical mixing across the contact interface between the two gasses, but
it also leads to more vigorous hydrodynamic mixing by resolving the turbulence
generated on smaller scales.  The result of these competing effects can be seen
in Figure \ref{fig:mixing}, where we compare the distribution of relativistic
gas in three 2D simulations with varying resolutions.   We also plot the effective
value of $R_p/R_1$ for these three simulations in
Figure \ref{fig:rpn2d}.  Up until
a time of $\sim 35,000$ yrs, all three simulations are quite similar.  Once
the Rayleigh-Taylor instability kicks in, however, the resolution-dependent
effects of mixing become quite evident.  
Independent of the numerical
resolution, it is clear that the mixing of relativisitc gas and thermal gas
is very efficient.

\subsection{Evolution in Nonuniform Media}

We have repeated our 2D simulation with the addition of a density
gradient in the ambient medium with the goal of understanding the displacement
of the pulsar bubble seen in the Vela SNR.
Following Dohm-Palmer \& Jones (1996), 
we use a smooth density distribution in the vertical direction given by
\begin{equation}
\rho (z) = \rho_c \left[ 1 + \frac{2-x}{x}\tanh (z/H)\right]
\end{equation}
such that the density contrast, from a maximum at large positive $z$ to a minimum
at large negative $z$, is given by $(x-1)^{-1}$, and $H$ is the characteristic
length scale over which the density changes.  The simulations shown here
use $x=1.2$ corresponding to a density contrast of 5.
These simulations require the computation of the full domain in $z$ (i.e., no
assumption of equatorial symmetry).  As a result, our standard grid of
2000 zones represents only half the resolution used in the previous 2D model.

\begin{figure}[!hbtp]
\plotone{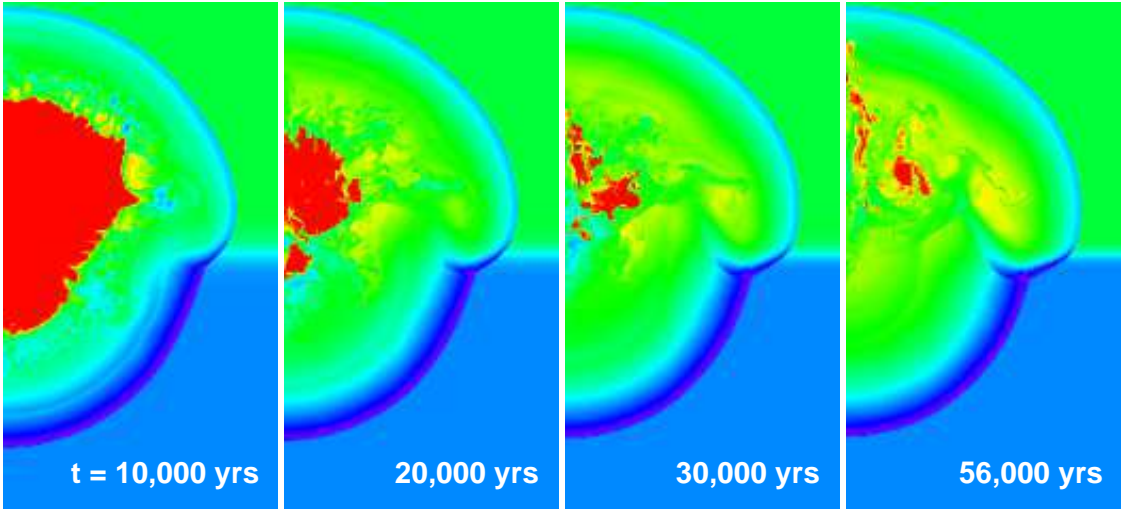}
\figcaption{The evolution of the pulsar nebula/supernova remnant for the
model with a scale height of $H=1\times 10^{19}$ cm in the ambient medium.
The logarithmic density scale covers 2.5 orders of magnitude.
\label{fig:varg1}}
\end{figure}

In Figure \ref{fig:varg1} we show the evolution of the PWN/SNR system for
an ambient medium length scale of $H=1\times 10^{19}$ cm, which corresponds 
to the size
of the SNR at an age of $\sim 500$ yrs.   
Early in the evolution, we can estimate the asymmetry in the SNR by assuming
only radial motion, such that the local radius is determined by
eq. (4) with the local preshock density.  The full density contrast of
5 thus produces a relatively minor ratio in radii at 
the poles of $5^{1/9}\approx 1.2$.  The ratio of pressures behind the
forward shocks is more pronounced: $5^{7/9}\approx 3.5$  Thus, while
the forward shock is only mildly asymmetric, the interior pressure
variation drives a substantial asymmetry in the process of crushing
the PWN. 

\begin{figure}[!hbtp]
\epsscale{0.9}
\plotone{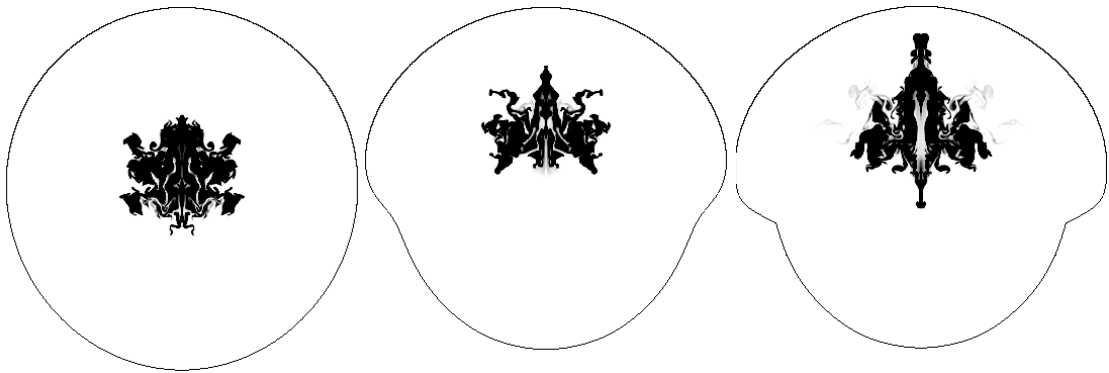}
\figcaption{The remnants of the pulsar nebula as defined by the
adiabatic index (black is gas with $\gamma < 5/3$) for simulations
with a scale height of $H=1\times 10^{20}$ cm (left) and
$H=3\times 10^{19}$ cm (center) and $H= 10^{19}$ cm (right).  The solid 
contours mark the location
of the shock front. These snapshots are from a time of $\sim 50,000$ years,
when the SNR has a radius of $\sim 1.5\times 10^{20}$ cm.
\label{fig:varg3}}
\end{figure}

If the ambient medium had an exponential density distribution instead
of a step in density, the distortion
of the outer shock front would not be as apparent (e.g., Hnatyk \&
Petruk 1999).
In Figure \ref{fig:varg3} we show the results of two additional simulations with
larger values of $H$, but all at the same age of $\sim$50,000 yrs.
In the simulation with $H=1\times 10^{20}$ cm, the asymmetry in the
CSM at the time of the PWN/SNR interaction is less than a factor of 2, 
and the crushed PWN is only slightly off center.  But a slightly steeper
density gradient, $H=3\times 10^{19}$ cm, begins to affect the SNR
prior to the PWN/SNR interaction, and the crushed PWN is displaced
from the center of the remnant by roughly $40\%$ the radius of the SNR.

\section{COMPARISON WITH OBSERVATIONS}

\subsection{The Vela Remnant}

A pulsar nebula that is likely to be in the post-reverse shock phase is 
Vela X in the Vela supernova remnant.
This nebula was first identified as a region with a flat radio spectral index
near the Vela pulsar (Weiler \& Panagia 1980).
More recent radio studies have shown that some of the emission is in filamentary
structures as well as a diffuse background (Milne 1995; Frail et al. 1997; Bock,
Turtle, \& Green 1998).
The extended Vela X emission is largely to the south of the pulsar.
The radio observations of Bock et al. (1998) give an especially detailed 
view of the Vela X nebula.
The overall appearance is chaotic, in keeping with the expected instabilities
during the crushing and re-expansion phases.
There are a number of radio filaments, and the magnetic field lines lie
along the filaments; this is expected if the Rayleigh-Taylor instability
occurs primarily across the magnetic field rather than along it.
In their Fig. 5, Bock et al. (1998) show the projected velocity of the
Vela pulsar, including the pulsar position 12,000 years ago;
it is near the North edge of the Vela X nebula.
Even if this age is incorrect, the direction of the pulsar velocity
cannot take it to the center of the nebula.
We propose that the reason for the asymmetry is an asymmetry in
the medium surrounding the supernova remnant.

The interstellar medium surrounding the Vela remnant is known to
be inhomogeneous.
From X-ray observations, Bocchino, Maggio, \& Sciortino (1999)
deduced that the
presence of various interstellar components, the lowest density
component having a hydrogen density $n_H\approx 0.06$ cm$^{-3}$.
From HI observations, Dubner et al. (1998) deduced a preshock
density $n_H\approx 1$ cm$^{-3}$; this denser gas is primarily on
the north side of the Vela remnant.
We thus propose that the supernova occurred at the 
pulsar position (corrected for its
velocity), that the higher density to North caused greater SNR
expansion to the south, and that the asymmetry in the reverse shock front
pushed the pulsar nebula to the south of the pulsar.
The age of the remnant can be estimated from the spindown age of the
pulsar, $t_{sd}=P/[(p-1)\dot P]$, where $p$ is the pulsar braking
index, $P$ is the pulsar period, and $\dot P$ is its time derivative.
The spindown age is the actual age if $p$ is constant, the pulsar magnetic
field is constant, and the current spin rate is much less than the initial
rate.
If this last assumption is violated, the age is $<t_{sd}$.
For the magnetic dipole model with $p=3$, we have
$t_{sd}=11,000$ years for the Vela pulsar 
(Reichley, Downs,  \& Morris 1970).
However, Lyne et al. (1996) found evidence for $p=1.4\pm 0.2$, 
which leads to a larger value of $t_{sd}\approx 56,000$ years.
These estimates show consistency with the hypothesis that the
reverse shock front has recently crushed the pulsar nebula in Vela.
Another expectation of the reverse shock model is that the pressure
in the pulsar nebula should be comparable to the pressure in the
hot gas in the Vela remnant.
Estimates of the pressure in the relativistic fluid from minimum
energy synchrotron emission arguments show that this is approximately
the case (Frail et al. 1997).

With the standard assumptions, the evolution of the pulsar spin rate is
\begin{equation}
\Omega = \Omega_i\left(1+{t\over\tau}\right)^{-1/(p-1)},
\end{equation}
where $\Omega_i$ is the initial pulsar spin rate and $\tau$ is
approximately the timescale for power input by the pulsar (see \S~3).
The present period of the Vela pulsar is $P=2\pi/\Omega=89$ ms.
If $p=1.4$ and $t\gg \tau$, we have $\Omega\approx
\Omega_i(t/\tau)^{-2.5}$ and the rapid evolution of the spin rate
requires a value of the initial spin rate that is faster than plausible
if $\tau\approx 500$ years.
This is true even for our model E with $\tau\approx 2500$ years.
However, the value of $p=1.4$ by Lyne et al. (1996) is uncertain
because of the frequent glitching of the pulsar and a value closer
to the magnetic dipole value of 3 is possible.
Of the models presented in \S~3, model A appears most suitable because
the radius of the pulsar nebula is 1/4 of the outer shock radius
after the crushing has occurred; the diameter of Vela X is $\sim 2^{\circ}$
(Bock et al. 1998), as compared to the $\sim 8^{\circ}$ diameter of
the Vela SNR.

The fact that the Vela pulsar is to one side of Vela X has led to
previous discussions of how power might flow from the pulsar into
the Vela X region.
Observations with {\it ROSAT} have shown an X-ray filament to the south of
the pulsar that has been interpreted as emission from a jet extending from the
pulsar (Markwardt \& \"Ogelman 1995).
It is not yet clear whether the X-ray emission is thermal or nonthermal
(Markwardt \& \"Ogelman 1997).
Frail et al. (1997) suggested that some of the radio structures are related to
the X-ray ``jet.''
However, the radio filament near the X-ray feature appears to be similar to
other radio filaments in the Vela X region (Bock et al. 1998).
Also, the X-ray spectrum of the feature is similar to that of the more diffuse 
X-ray emission
in the Vela X region (Markwardt \& \"Ogelman 1997).
Wijers \& Sigurdsson (1997) noted difficulties with the high momentum
in the proposed X-ray jet and suggested a different model combining
fallback onto the neutron star with a trail left by a 
high velocity binary companion.

The model described here suggests a different interpretation of 
the ``jet'' features.
As mentioned above, the linear feature in the radio synchrotron
emission can be the result of the Rayleigh-Taylor instability
during the crushing process.
The instability also results in hot, thermal gas being mixed into
the region initially occupied by the relativistic fluid.
Because of the sweeping action of the pulsar nebula, the gas
immediately outside the pulsar nebula has a higher density than
that occupying most of the SNR (Figs. 1 and 2).
It is thus plausible that some of the gas mixed into the
pulsar nebula has a higher emissivity and appears as an X-ray
filament.
Frail et al. (1997) found that the radio and X-ray emission are
not coincident but lie near each other, as would be expected in
our model.

\subsection{Other Remnants}

The return of a reverse shock front to the center of a supernova
remnant is not generally expected to  be spherically symmetric
because of inhomogeneities in the surrounding medium.
Thus we might expect our models to be broadly applicable to
composite supernova remnants which have both an apparent pulsar
nebula and a well-developed shell (see Helfand 1998 for a list of objects).
The radio emitting electrons are long lived and are the best
tracer of the power input from the pulsar during its phase
of high power output.
The X-ray emitting electrons are shorter lived and provide
an indication of the current position of the pulsar, even if
a pulsar has not been directly observed.
However, an observable shell may be present before the reverse
shock crushing phase.
This appears to be the case, for example, in the Large Magellanic
Cloud remnant 0540--69, which has a pulsar nebula very similar to
the Crab Nebula (Kaaret et al. 2001).
At late times, the mixing of the radio emitting electrons with
the thermal gas may lead to the disappearance of the radio nebula
from the initial injection phase.
In this case, only emission from near the pulsar, likely to be in
a trail because of the pulsar velocity, is present.
 
In the Vela remnant, the pulsar is surrounded by an X-ray PWN,
while the radio emitting PWN is displaced from the X-ray emission.
A similar situation appears to be present in the remnants
MSH15--56 (G326.3--1.8; Plucinsky 1998; Gaensler 2001),
G327.1--1.1 (Sun, Wang, \& Chen 1999), and G0.9+0.1
(Gaensler, Pivovaroff, \& Garmire 2001) although pulsars associated
with the X-ray emission have not yet been discovered in these cases.
The broad regions of radio emission are   excellent candidates
for wind nebulae injected early in the pulsar evolution and
crushed by the reverse shock.

A less clearcut case is the PWN inside W44 (Frail et al. 1996;
Harrus et al. 1996).
The radio nebula has a bow shock appearance, but the trailing
emission is considerably broader and stronger than that near
the pulsar.
The ratio of PWN radius to outer shock radius is $\sim 0.1$.
Although smaller than the case of Vela, this situation is expected
in our models (see Fig. 3).
The age of W44 is $\sim 20,000-30,000$ years, but the relatively
high density of its surroundings places it in a more evolved
category than Vela (Chevalier 1999).
Another remnant with relatively dense surroundings is IC 443,
with an estimated age of 30,000 years (Chevalier 1999).
Early radio spectral observations indicated the presence of
a PWN (Green 1986), and recent X-ray and radio observations clearly
show a bow shock nebula with a compact X-ray source at the head
(Olbert et al. 2001).
In this case, the radio emission region is narrow and is strongest
close to the compact X-ray source; there is no clear sign of
emission from a remnant nebula that has been crushed by the
reverse shock front.
The trailing emission is an approximately linear feature, but does
not point toward the center of the remnant.
This appears to be another case of an off-center explosion due to
 asymmetries in the surrounding medium.
This situation may account for some of the high velocities deduced
for pulsars from their positions relative to what appears to be
their parent SNR (Frail, Goss, \& Whiteoak 1994).

\section{DISCUSSION AND CONCLUSIONS}

We have considered the evolution of PWN (pulsar wind nebulae) inside of
supernova remnants, with special attention to the effect of the
reverse shock front on the PWN.
It is possible that synchrotron losses are important for the
PWN at early times, but they are unlikely to be significant at the
times of interest here and the effect can be approximately accounted
for by renormalizing the pulsar spin down energy.
We do not treat the instabilities that can occur early in the
evolution as a result of the acceleration of the shell outside the
pulsar nebula (Jun 1998); these would only further contribute
to the instabilities that we find later in the evolution.

Our simulations allow for two fluids: a $\gamma=4/3$ fluid for the
PWN and a $\gamma=5/3$ fluid for the surrounding gas.
We do not allow for the magnetic field that is expected to be
mixed in with the relativistic particles and that can influence
the evolution through magnetic tension.
The field might affect the Rayleigh-Taylor instabilities that occur
when the PWN is compressed by the reverse shock gas and subsequently re-expands.
The magnetic field can dampen the instability along magnetic field
lines, but not across them, leading to the creation of filaments along
the magnetic field direction.
There is mixing between the fluids so that the separation of the
two fluids depends on the numerical resolution of the two-dimensional
simulations.
The degree of mixing may also depend on the magnetic field.

In addition to causing mixing of thermal gas with the PWN, the reverse
shock wave is unlikely to be spherically symmetric and can thus
push the PWN off center from the explosion site.
The position of a pulsar can be displaced from the explosion site
both because of this effect and because of the velocity that
the pulsar acquires at birth.

In comparing the model results with observations, it is important
to correctly identify the evolutionary state of a PWN.
During the initial phase when the PWN is interacting with
the freely expanding supernova ejecta, the pulsar is expected
to be approximately centrally located within the PWN.
Even if the pulsar has a significant velocity, it will not have
a particular position in the uniformly expanding medium unless there
is a large density gradient in the medium.
Nebulae likely to be in this phase include the Crab Nebula and 3C 58.

The next event in the evolution is the cessation of power input
from the central pulsar.
The outer edge of the nebula continues its free expansion, leading
to a low surface brightness nebula that can have a substantial 
filling fraction within the external shock wave.
The nebula MSH 15-52 around the pulsar PSR B1509--58
(Seward et al. 1984) may be in this phase.

Here, we have emphasized the late evolutionary phase after the
reverse shock front comes back to the center of the remnant.
An excellent candidate for this phase of evolution is the Vela X
radio remnant within the Vela supernova remnant.
The fact that the pulsar nebula is displaced from the pulsar can
be explained in our model without the need for directed power
from the pulsar.
Asymmetries in the placement of radio PWNae relative to their
pulsars are likely to be common in this phase because the
interstellar medium is generally inhomogeneous.

\acknowledgments
Support for this work was provided in part by NASA grants NAG5-8130
and NAG-7153.  The hydrodynamic simulations were computed on an IBM
SP at the North Carolina Supercomputing Center.

\end{document}